%Paper: hep-th/9210079
%From: bars@physics.usc.edu (Itzhak Bars)
%Date: Wed, 14 Oct 92 19:23:47 PDT
%Date (revised): Fri, 30 Oct 92 14:26:46 PST

%%%%%%%%%%%%%%%%%%%%%%%%%%%%%%%%%%%%%%%%%%%%%%%%%%%%%%%%%%%%%%%%%%%%%

% Lecture at Erice 1992 and Salamanca 1992

%%%%%%%%%%%%%%%%%%%%%%%%%%%%%%%%%%%%%%%%%%%%%%
\input harvmac

\baselineskip=13pt
\overfullrule=0pt
%%%%%%%%%%%%%%%%%%%%%%%%%%%%%%%%%%%%%%%%%%%%%%%%%%%%%%%%%%%%%%%%

\def\a{\alpha}

\def\b{\beta}

\def\g{\gamma}

\def\d{\delta}
\def\D{\Delta}
\def\e{\epsilon}

\def\m{\mu}
\def\n{\nu}

\def\no{\noindent}
\def\hb{\hfill\break}

\def\qq{\qquad}

\def\bl{\bigl}
\def\br{\bigr}

%%%%%%%%%%%%%%%%%%%%%%%%%%%%%%%%%%%%%%%%%%%%%%%%%%%%%%%%%%%%%%%%%

\def\IR{\relax{\rm I\kern-.18em R}}

%\nopagenumbers
%%%%%%%%%%%%%%%%%%%%%%%%%%%%%%%%%%%%%%%%%%%%%%%%%%%%%%%%%%%%%%

hep-th/9210079  \hfill  {USC-92/HEP-B5}
\rightline {Oct. 1992}
\bigskip\bigskip

\centerline {  {\bf SUPERSTRINGS ON CURVED SPACETIMES}
 \footnote{$^*$}{Lecture delivered at the International Workshop on ``String
Quantum Gravity and Physics at the Planck Scale", Erice, Italy, June 1992,
and \hb
XIX$^{th}$ International Colloq. on Group Theoretical Methods in Physics,
Salamanca, Spain, June 1992.}
 \footnote{$^\dagger$}{Research supported in part by DOE under Grant No.DE-
FG03-84ER-40168.} }
 %\centerline {\bf AND NONCOMPACT GROUPS}

 \bigskip\bigskip
 \centerline {ITZHAK BARS}

 \bigskip
 \centerline {Physics Department, University
of Southern California}
 \centerline {Los Angeles, CA 90089-0484, USA}
 \bigskip\bigskip\bigskip

\centerline {ABSTRACT}

\bigskip

In this lecture I summarize recent developments on strings propagating in
curved spacetime. Exact conformal field theories that describe gravitational
backgrounds such as black holes and more intricate gravitational singularities
have been discovered and investigated at the classical and quantum level.
These models are described by gauged Wess-Zumino-Witten models, or
equivalently current algebra G/H coset models based on non-compact groups,
with a single time coordinate. The classification of such models for all
dimensions is complete. Furthermore the heterotic superstrings in curved
spacetime based on non-compact groups have also been constructed.
For many of the $d\le 4$ models the gravitational
geometry described by a sigma model has been determined. Some general results
outlined here include a global analysis of the geometry and the exact
classical geodesics for any G/H model. Moreover, in the quantized theory, the
conformally exact metric and dilaton are  obtained for all orders in an
expansion of $k$ (the central extension).
All such models have large-small (or mirror) duality properties which we
reformulate as an inversion in group space.
To illustrate model building techniques a specific 4-dimensional
heterotic string in curved spacetime is presented. Finally the methods for
investigating the quantum theory are outlined. The construction and analysis
of these models at the classical and quantum level involve some aspects of
noncompact groups which are not yet sufficiently well understood. Some of the
open problems in the physics and mathematics areas are outlined.

\vfill\eject
%%%%%%%%%%%%%%%%%%%%%%%%%%%%%%%%%%%%%%%%%%%%%%%%%%%%

\newsec {Introduction}

I would like to begin this talk by explaining why non-compact current algebra
coset models, or equivalently gauged Wess-Zumino-Witten (WZW) models based on
certain non-compact groups, are of great interest in string theory on curved
spacetime.

The first quantized version of string theory envisages a general sigma
model-like action that describes a string $X^\mu(\tau,\sigma)$ propagating in
a general background that includes a metric $G_{\mu\nu}(X)$, dilaton
$\Phi(X)$, antisymmetric field $B_{\mu\nu}(X)$, tachyon $T(X)$, as well as all
fields representing all other massive modes of the string. These background
fields are imagined to be determined dynamically in the fully interacting
string theory, including non-perturbative effects. Conformal invariance is a
``kinematical" property of string theory. However, it is such a strong
constraint that it generates dynamical equations for these background fields.
One may hope that a solution to these dynamical equations provides a
``classical vacuum" to string theory that captures some of the important
dynamical properties of the theory. If one could really determine the vacuum
state one would hope that it explains all the essential properties of low
energy physics, including the facts that we live in four dimensions ($D=4$),
that the color-electroweak gauge group is $SU(3)\times SU(2)\times U(1)$, that
there are three families of quarks and leptons, etc.. In addition, one may
derive a quantum cosmological history of the universe and learn about the
mechanism for mass generation. If all this is accomplished new predictions
should follow.

In the past the main tool for imposing conformal invariance in string theory
in curved spacetime has been the beta function conditions of
Fradkin and Tseytlin and Callan et. al. These conditions have the form of
Einstein's equations modified by the presence of a dilaton and some additional
fields such as an antisymmetric tensor, etc. In fact they are identical to the
equations of motion of the low energy (perturbative) effective action of
string theory. These equations represent only a perturbative version of string
theory in two senses: (i) they apply only in the absence of string loops (i.e.
they apply only on the topology of the sphere, not the torus, pretzel, etc.)
and (ii) the non-linear sigma model interactions are considered only in lowest
orders. There is not much we can do about (i) at this stage (except for the
non-perturbative matrix model type of treatment of two dimensional gravity);
one maintains the hope that the string loop expansion is useful to extract at
least part of the physics, in the same sense that perturbative expansions have
been useful in field theories such as QED and QCD. On the other hand there is
much we can and should do to improve (ii) since the basic symmetry of
conformal invariance is satisfied only perturbatively by the solutions of the
beta function equations.

Fortunately, conformal invariance is exactly satisfied in a class of models
based on current algebra cosets or equivalently gauged WZW
models (still excluding string loops). The beta function equations are then
automatically solved in the perturbative limit of the model. But more
significantly, conformal invariance is satisfied non-perturbatively to all
orders of the central extension $k$ (equivalently to all orders of the non-
linear sigma model interactions). When a non-compact group is used there are
automatically both spacelike as well as timelike coordinates. Only one
timelike coordinate can be tolerated since, roughly, the conformal constraints
(Virasoro constraints) are sufficient to remove the negative norm states
generated by only one timelike coordinate - this is analogous to the naive
counting that leads to the no-ghost theorem in flat spacetime. To achieve the
single time requirement
 \ref\BN{I. Bars and D. Nemeshansky, Nucl. Phys. {\bf B348} (1991) 89. }
one must restrict the possible non-compact groups and their cosets to the
following list
\ref\IBCS{I. Bars, ``Curved Space-Time
Strings and Black Holes", in Proc. of {\it XX$^{th}$ Int. Conf. on Diff.
Geometrical Methods in Physics}, Eds. S. Catto and A. Rocha, Vol.2, p.695,
(World Scientific, 1992).}
\foot{The star $*$ in the last two lines of $(1.1)$ means that one must take
the non-compact version of $G$ such that the maximal compact subgroup is
$H\times U(1)$, and $H$ is the subgroup that appears in the denominator of
$G^*/H$. } :

\eqn\list{\eqalign {
  SO(d-1,2)/SO(d-1,1) \qquad & SO(d,1)/SO(d-1,1) \cr
  SU(n,m)/SU(n)\times SU(m) \qquad & SO(n,2)/SO(n) \cr
  SO(2n)^*/SU(n)        \qquad &  Sp(2n)^*/SU(n)    \cr
  E_6^*/SO(10)         \qquad  &  E_7^*/E_6 } }
This list, which contains only simple groups, may be extended with direct
products of simple groups $G_1\times G_2\times \cdots$ including $U(1)$ or
$\IR$ factors, or their cosets, so long as the additional factors do not
introduce additional time coordinates \BN\IBCS
 \ref\GIN{ P. Ginsparg and F. Quevedo, ``Strings on Curved Space-Times
 Black Holes, Torsion, and Duality'', LA-UR-92-640.}.
While these models represent only a small subset of all possible curved
spacetime models described by the general sigma model, they have the advantage
of being solvable in principle thanks to the current algebra formulation. Thus
a lot more can be said about the spectrum, correlation functions, etc. of the
quantum string theory based on these models. Furthermore, it has been realized
that the special geometries described by these non-compact groups are relevant
to gravitational singularities such as black holes and cosmological Big Bang.
For these reasons this class of models has received considerable attention
during the past year and a half
 \ref\WIT{E. Witten, Phys. Rev. {\bf D44} (1991) 314. }
 \ref\IBSFthree{I. Bars and K. Sfetsos, Mod. Phys. Lett. {\bf A7} (1992)
1091.}
 \ref\others{
 M. Crescimanno. Mod. Phys. Lett. {\bf A7} (1992) 489. \hb
 J. B. Horne and G.T. Horowitz, Nucl. Phys. {\bf B368} (1992) 444. \hb
 E. S. Fradkin and V. Ya. Linetsky, Phys. Lett. {\bf 277B} (1992) 73. \hb
 P. Horava, Phys. Lett. {\bf 278B} (1992) 101.\hb
 D. Gershon, TAUP-1937-91 \hb }.
It is hoped that through such solvable models new light will be shed on
unresolved gravitational issues, in string theory as well as general
relativity, such as  singularities, quantization and finiteness or
renormalizability in curved spacetime, the question of Euclidean-Minkowski
continuation, spectrum of low energy particles and excited string states in
the presence of curvature, etc..

The other important question for string theory is the nature and content of
the low energy matter it is supposed to predict in the form of quarks, leptons,
gauge bosons, etc.. The new models have opened up the possibility of heterotic
superstring theories in four spacetime
dimensions (with or without additional compactified dimensions)
 \ref\IBSFhet{I. Bars and K. Sfetsos, Phys. Lett. {\bf 277B} (1992) 269.}.
This
is possible because the $c=26$ (or $c=15$ with supersymmetry) condition
can be satisfied in fewer dimensions provided the space is curved. For
example it has been possible to construct consistent purely four dimensional
heterotic string theories based on non-compact current algebra cosets
 \ref\IBhe{I. Bars, Nucl. Phys. {\bf B334} (1990) 125. }
 as will be illustrated in the next section. The gauge groups that emerge fall
within a remarkably narrow range and include the desirable low energy color-
electroweak symmetry of $SU(3)\times SU(2)\times U(1)$. The quark and lepton
states, which come in color triplets and $SU(2)$ doublets, are expected to
emerge in several families. Compared to the popular approach of four flat
dimensions plus compactified dimensions, the gauge groups are either the same
or closely related. This gives the hope that the spectrum of a curved purely
four dimensional heterotic superstring that describes the very early universe
may be closely related to the quarks and leptons that survive to the present
times.

 Thus, it is of interest to investigate this spectrum for a purely four
dimensional heterotic superstring theory in curved spacetime with or without
additional compactified dimensions. One advantage of
pure four dimensions is the expected tighter predictions since the multitude
of string ``vacuua" associated with the higher compactified dimensions would
be avoided. This approach does not explain the deeper issue of why there are
four dimensions in the physical world, but allows the exploration of the kinds
of results that would emerge if the Universe has in fact only four dimensions
at all times
 \foot{From the point of view of perturbative conformal conditions there seems
to be a problem with the incompatibility of Poincar\'e invariance and $c=26$
in purely four dimensions (e.g. $c=26$ is possible for an asymptotically flat
spacetime in four dimensions provided the dilaton is asymptotically linear).
This problem may or may not be resolved with a better understanding of
conformal invariance and non-perturbative effects such as phase transitions
and the relation of the observed physical flat universe to the early curved
string universe. Presumably the curved spacetime string theory {\it must}
undergo some physical phase transitions, including giving a mass to the
dilaton and the usual inflationary scenario, before it can be connected to the
observed flat, homogeneos and isotropic Universe. In the final stage one
expects an effective field theory of the massless particles that decribes the
low energy physics below the Planck scale. This is no longer the full string
theory, and it is at this stage that one requires a flat, homogeneous and
isotropic universe after inflating a small part of the early universe into our
present Universe. Thus, there may be a resolution for conformal invariance on
the one hand and four flat dimensions on the other, after taking into
consideration  the physical situation. If this proves to be impossible then
one must accept more than four dimensions and consider the non-compact group
models in higher dimensions and/or  direct products with an internal conformal
field theory. The methods for computing the low energy spectrum are basically
the same in a model with either pure four dimensions or with higher
compactified dimensions included.}.

The study of specific four dimensional manifolds that emerge in these models
has proceeded through some rather special group theoretical methods. Some of
it should be of interest in classical gravitational physics and would seem
rather remarkable to general relativists. For example, it has been possible to
determine the global properties of the singular geometries and solve
completely for all particle geodesics
 \ref\IBSFglobal{I. Bars and K. Sfetsos, ``Global Analysis of New
Gravitational Singularities in String and Particle Theories", USC-92/HEP-B2
(hep-th/9205037), to appear in Phys. Rev. D (1992)}
 \ref\IBSFslsu{I. Bars and K. Sfetsos, ``$SL(2\IR)\times SU(2)/\IR^2$ String
Model in Curved Spacetime and Exact Conformal Results", USC-92/HEP-B3 (hep-
th/9208001), to appear in Phys. Lett. }.
These global manifolds contain several copies of the same
``world" and geodesics continue smoothly from one world to the next by going
through curvature singularities. The general solution to {\it string
geodesics} has also been outlined
\IBSFthree\IBSFglobal .

The quantum investigations have also gone well beyond the one loop
approximations. For example, it has been possible to determine the conformally
exact metric and dilaton to all orders in the sigma model interactions
 \ref\IBSFexa{I. Bars and K. Sfetsos, ``Conformally Exact Metric and Dilaton
in String Theory on Curved Spacetime", USC-92/HEP-B2 (hep-th/9206006), to
appear in Phys. Rev. D (Nov. 1992)}.
The new algebraic methods, which will be described below apply to all gauged
WZW models.

\newsec{ Curved Spacetime and Non-compact Groups}

\subsec{Time Coordinate}

Let us consider a WZW model based on a non-compact group. Let us parametrize
the group element by $X^A(\tau,\sigma)$, where $A$ is an index in the adjoint
representation. The left or right moving currents take the form $J^A=\partial
X^A+\cdots$, where the dots stand for non-linear terms in an expansion in
powers of $X$. The Fourrier components of these currents $J_n^A$ satisfy a
Kac-Moody algebra

\eqn\kacmoody{ [J_n^A,J_m^B]=if^{AB}_C J_{n+m}^C - k{n\over 2}\eta^{AB}
\delta_{n+m,0} }
where $k$ is the central extension and $\eta^{AB}$ is proportional to the
Killing metric. In an appropriate basis one can choose a diagonal $\eta^{AB}=
diag (1,\cdots,1, -1,\cdots -1)$ with $+1$ entries corresponding to compact
generators and $-1$ entries to non-compact ones. For example, for $SL(2,\IR)$
with currents $(J^0,J^1,J^2)$, one has the Minkowski metric in $2+1$
dimensions: $\eta^{AB}=diag (1,-1,-1)$.

Let us consider the large positive $k$ limit of the WZW model and examine
the commutation rules of its canonical currents. It is convenient to
define the rescaled currents $\alpha_n^A=\sqrt{2\over k}J_n^A$. When
$k\to\infty$ these behave like the free field oscillators of the flat string
theory (either left or right movers)

\eqn\free{ [\alpha_n^A,\alpha_m^B]= -n\eta^{AB}\delta_{n+m,0}. }
We see that, in the large $k$ limit, we have free field degrees of freedom
$X^A\sim \sum_n {1\over n}\alpha_n^A z^n + \cdots $ ,that behave like time
coordinates when $A$ corresponds to compact generators and like space
coordinates when $A$ corresponds to  non-compact generators. The signature of
the coordinates are the same for finite positive $k$. This is seen by
specializing the commutation rules \kacmoody\ to $A=B$ for which the structure
constant of the Lie algebra drops out.

In a string theory one can tolerate only one time coordinate. This is because,
by naive counting, the Virasoro constraints $L_n\sim 0$ can eliminate only
the ghosts generated by the negative norm of one time-like oscillator
$\alpha_n^0$, just like string theory in flat spacetime. Therefore, one must
put constraints that set to zero the unwanted compact generators, except for
one of them. However, first class constraints must close to form an algebra.
Therefore, the currents that are set equal to zero ($J^a\sim 0$ weakly on
states) must form a subalgebra corresponding to a subgroup of the non-compact
group $H\in G$. The subalgebra may include compact and non-compact generators.
The remaining currents $J^\mu, \ \mu=0,1,2,\cdots (d-1)$ stand in one-to-one
correspondance to the coset coordinates $X^\mu$ that include just one time
coordinate. Thus, one must choose a subgroup $H$ such that the coset $G/H$ has
the signature of Minkowski space in $d$ dimensions. It is well known that this
set of constraints defines an exact conformal field theory that fits the
algebraic framework of GKO.
The new ingredient is that one must take an appropriate non-compact coset
$G/H$. The only simple groups that give a single time coordinate were
classified in \IBCS\ and are listed in \list .

There is another way to see the same result by using a Lagrangian method
at the classical level rather than the algebraic Hamiltonian argument given
above at the quantum level. A GKO theory correponds to a
gauged WZW model with the subgroup $H$ local. Using the gauge invariance one
can eat-up $dim(H)$ degrees of freedom, leaving behind $dim(G/H)$ group
parameters that contain just one timelike coordinate. Since the gauge fields
are non-dynamical they can be integrated out. This leaves behind a sigma model
type theory with the desired signature. The large $k$ limit of this theory
has free field quantum oscillators with a single time coordinate.

Both the Hamiltonian and Lagrangian arguments were first given by Bars and
Nemeschansky \BN . The Hamiltonian approach was given more weight in \BN\
where several examples, including $SL(2,\IR)/\IR$ at $k=9/4$, were
investigated. The Lagrangian method was explicitly carried out for
$SL(2,\IR)/\IR$ by Witten \WIT
 who interpreted the sigma model metric as a black hole. With the realization
that non-compact group coset methods generate singular geometries there has
been a flurry of activity to determine the geometries of higher dimensional
cosets \IBSFthree\others\IBSFhet\GIN\
as discussed in the following sections.

\subsec{Action for Heterotic Superstring in Curved Spacetime}

There are additional ingredients that must be included in a physical model. A
good model must not have a tachyon. The large $k$ limit that reduces to a flat
string theory provides a guide for how to eliminate the tachyon state. Namely,
one must start with a string theory that has at least $N=1$ supersymmetry on
the world sheet, and then impose the GSO projection
\ref\GSO{F. Gliozzi, J. Scherk and D. Olive, Nucl. Phys. B122 (1977) 253.
 \semi M. B. Green, J. H. Schwarz and E. Witten, {\it Superstring Theory},
Vol.1, p.218, (Cambridge 1987) }
on the spectrum $(-1)^F=1$. This is achieved for any $k$ in curved spacetime
by starting from a Kazama-Suzuki type model based on non-compact cosets
\IBhe . Thus, the super coset may be given in the form

\eqn\supercoset{ {G_{-k}\times SO^*(dim(G/H))_1\over H_{-k+g-h} } }
where $SO^*$ is a non-compact version of $SO$.
Moreover, a physical model must be a heterotic theory that includes gauge
groups. This is done by taking the coset \supercoset\ with central charge
$c=15$ for left movers and the coset $G_{-k}/H_{-k}\times (gauge\  group)$
with $c=26$ for right movers. We can then search for all possibilities that
satisfy these requirements. In what follows, for definiteness, we
restrict ourselves to cosets $G/H$ that have only four bosonic dimensions. It
is evident that this restriction is not a priori justified in our formalism
and evidently more general models with higher dimensions are possible.
However, as emphasized above it is of interest to find out the behaviour of
purely four dimensional models of this type (see footnote 2).

At this point let us construct the gauged WZW action for a heterotic
superstring in curved spacetime. Here we first repeat the
$SO(3,2)/SO(3,1)$ example worked out in \IBSFhet
and emphasize
a few important points.
In the
conformal gauge the action has
four parts $S=S_0+S_1+S_2+S_3$ with

 \eqn\action{ \eqalign {
 &S_0(g)={k\over 8\pi}\int_M d^2\sigma\ Tr(g^{-1}\partial_+g\
g^{-1}\partial_-g)
 -{k\over 24\pi}\int_B Tr(g^{-1}dg\ g^{-1}dg\ g^{-1}dg) \cr
 &S_1(g,A)=-{k\over 4\pi}\int_M d^2\sigma\ Tr(A_-\partial_+gg^{-1}
-\tilde A_+g^{-1}\partial_-g + A_-g \tilde A_+g^{-1}-A_-A_+)\cr
 &S_2(\psi_+ ,A_-)=-{k\over 4\pi}\int_M d^2\sigma\ \psi_+^\mu (iD_-
\psi_+)^\nu  \eta_{\mu\nu}, \qquad S_3(\chi_-)= {k\over 4\pi}\int_M d^2\sigma\
\sum_{a=1}^{22}\ \chi_-^a i\partial_+\chi_-^a
 }}
In addition, there are ghost actions $S_4(b_{L},c_{L},\beta_{L},\gamma_L)$
for left movers and $S_5(b_R,c_R)$ for right movers that are added due to the
superconformal or conformal gauge fixing respectively. This action has $(1,0)$
superconformal symmetry (see below) and is appropriate for the heterotic
string. The type-II string requires $(1,1)$ superconformal symmetry. Its
action follows if $\chi_-^a$ is removed and replaced by $\psi_-^\mu$ that
appears with a gauge covariant kinetic term just like $\psi_+^\mu$. Then
$S_3,\ S_5$ are replaced by $S_3(\psi_-,A_+)$ and
$S_5(b_R,c_R,\beta_R,\gamma_R)$.

In the above, $S_0$ is the global WZW model
 \ref\WITT{E. Witten, Comm. Math. Phys. 92 (1984) 455. }
with $g(\sigma^+,\sigma^-)\in SO(3,2)$. By itself this piece has
$SO(3,2)_L\times SO(3,2)_R$ symmetry. Since $SO(3,2)$ has a non-Abelian
compact subgroup $SO(3)$ the quantum path integral could be defined uniquely
only for $k=integer$ (this was not a restriction for $d=2,3$).
\foot{ The easiest way to see this point is to write $g$ in parametric form
$g=abc$ with $a\in SO(3)\ , b\in SO(2)$ and $c\in SO(3,2)/SO(3)\times SO(2)$
and apply the Polyakov-Wiegman formula
\ref\PW{A. M. Polyakov and P. W. Wiegman, Phys. Lett. B131 (1983) 121. }.
Then $S_0(g)$ decomposes into several pieces one of which is $S_0(a)$ that
can be defined only for integer $k$ since $SO(3)$ is compact \WITT . The
remaining pieces do not present a problem. }
Indeed, we take $k=5$ which is the value required by the {\it total} Virasoro
central charge for the supersymmetric left movers \BN

\eqn\cc{ c_L={3kd\over 2(k-d+1)}=15 \qquad {\rm for} \qquad d=4,\quad k=5 . }
$c_L$ is cancelled by the super ghost system of $S_4$.
For type-II the central charge of the supersymmetric right-movers
is also $c_R=15$. However, for the heterotic string the bosonic part
$SO(3,2)_{-k}/SO(3,1)_{-k}$ gives

\eqn\cright{ c_R(bose)={10 k\over k-3}-{6k\over k-2} = 15  }
for the special value $k=5$ (already fixed in the action). Since the ghosts in
$S_5(b_R,c_R)$ contribute $-26$ we require a $c_R(\chi)=11$ contribution from
the free fermions $\chi_-^a$. Therefore the action $S_3$ contains $22$ free
fermions. This action could be viewed as giving rise to $SO(22)_1$ current
algebra theory for right movers. There are many other ways of obtaining
$c_R=11$ as exact conformal theories based on current algebras.
\foot{Some examples are $[(E_8)_1\times SU(4)_1]$, $[(E_7)_1\times SU(5)_1]$,
$[(E_7)_1\times SU(3)_1\times SU(2)_1\times U(1)]$, $[(E_6)_1\times
SO(10)_1]$, $[(E_6)_1\times SU(4)_2]$, $[SO(10)_2\times SU(3)_1]$, etc. }.

The second piece in the action $S_1$ gauges
 \ref\WZW{E. Witten, Nucl. Phys. B223 (1983) 422.
 \semi K. Bardakci, E. Rabinovici and B. Saering, Nucl. Phys. B301
(1988) 151.
 \semi K. Gawedzki and A. Kupiainen, Nucl. Phys. B320 (1989) 625.
 \semi H.J. Schnitzer, Nucl. Phys. B324 (1989) 412.
  \semi D. Karabali, Q-Han Park, H.J. Schnitzer and Z. Yang, Phys. Lett. B216
(1989) 307.
 \semi D. Karabali and H.J. Schnitzer, Nucl. Phys. B329 (1990) 649.}
the Lorentz subgroup $H=SO(3,1)$ which is embedded in $SO(3,2)_L\times
SO(3,2)_R$ with a deformation. As explained in \IBSFthree\ the action of the
gauge
group could be deformed on the left or the right of the group element $g$. If
the matrix representation of the gauged Lorentz algebra on the left is $t_a$
and the one on the right is $\tilde t_a$ then gauge invariance is satisfied by
$\tilde t_a=g_0^{-1}t_ag_0$ or $\tilde t_a=g_0^{-1}(-t_a)^Tg_0$, where $g_0$
is any constant group element in {\it complexified} $SO(3,2)$ (including
$g_0$'s not continously connected to the identity) and $t^T$ is the transpose
of the matrix. In this notation the action $S_1$ is expressed in terms of
$A_{\pm}=A_{\pm}^at_a$ and $\tilde A_{\pm}= A_{\pm}^a\tilde t_a$ with the same
$SO(3,1)$ gauge potential $A_\pm^a(\sigma^+,\sigma^-)$. The simplest case of
$\tilde t_a=t_a$ corresponds to the standard vector subgroup. The remaining
cases generalize the vector/axial gauging options that were first noticed for
the 2d black hole
\ref\GIV{A. Giveon, ``Target Space Duality and Stringy Black Holes", LBL-30671}
 \ref\IBBH{I. Bars,``String Propagation on Black Holes'', USC-91/HEP-B3.}
 \ref\DVV{R. Dijkgraaf, H. Verlinde and E. Verlinde, ``String Propagation in a
Black Hole Background", PUPT-1252 (1991)}
 \ref\KIR{E. Kiritsis, ``Duality in Gauged WZW Models'', LBL-30747.}
and thus provide a generalization of the concept of duality.
Examples are given in \IBSFthree .

The action $S_2$ contains the fermions $\psi_+^\mu $ with $\mu=0,1,2,3$ that
belongs to the coset $SO(3,2)/SO(3,1)$. The flat Minkowski metric
$\eta_{\mu\nu}=diag\ (1,-1,-1,-1)$ is used to contract the Lorentz indices. As
shown in
 \ref\KS{Y. Kazama and H. Suzuki, Nucl. Phys. B321 (1989) 232. }\
coset fermions lead to $N=1$ superconformal symmetry. Indeed the super coset
scheme $SO(3,2)_{-5}\times SO(3,1)_1/SO(3,1)_{-4}$ for left movers
requires that they appear with gauge covariant derivatives $D_-
\psi_+^\mu=\partial_-\psi_+^\mu-(A_-)^\mu{}_\nu\psi_+^\nu$. The explicit
supersymmetry transformations are written more conveniently in terms of the
$5\times 5$ matrix $\psi_+=\left (\matrix {0 & -\psi_+^\nu \cr \psi_{+\mu} &
0\cr }\right )$ that belongs to the $G/H$ part of the Lie algebra
\ref\WITKS{ E. Witten, ``The N Matrix Model and Gauged WZW Models", IASSNS-
HEP-91/26 .}.
\eqn\susy { \delta g=i\epsilon_-\psi_+g , \qquad
 \delta\psi_+=\epsilon_- (gD_+g^{-1})_{G/H}, \qquad \delta \chi_-^a=0,
\qquad  \delta A_\pm=0 , }
with  $\partial_-\epsilon_-(\sigma^+)=0$. In a type-II theory $\psi_-^\mu$
also mixes under  supersymmetry with the group element $g^{-1}$ with a
transformation similar to the one above. The independent right-moving
supersymmetry parameter in this case is $\epsilon_+(\sigma^-)$.

This theory is supplemented with the original GSO projection \GSO\
adapted to four dimensions. Namely, we construct the operator $(-1)^F$ with
the same prescriptions as \GSO\ and project onto the states $(-1)^F=1$. Let us
describe the effect on the ground states in the Neveu-Schwarz and Ramond
sectors for left movers. In our coset scheme these are conveniently labelled
by the scalar, vector and the two spinor representations of the fermionic
$SO(3,1)_1$. The GSO projection eliminates the scalar and one of the spinor
representations so that the tachyon is eliminated from the theory. The
remaining vector and Weyl spinor form the representations $({1\over 2},{1\over
2})$ and $({1\over 2},0)$ of the Lorentz group in four dimensions. As is well
known this is a covariant space-time supersymmetric vector multiplet and
therefore signals the possibility of space-time supersymmetry in our heterotic
model. The GSO projections for the type-II theory can be chosen such that the
remaining ground state Weyl spinors for the left movers and right movers have
either the opposite or the same chirality. Accordingly the theory will be
called type-IIA or type-IIB respectively. The important aspect of the GSO
projection is to eliminate the tachyon. After the projection one is
automatically left over with an equal number of fermionic and bosonic states.
It is not clear whether these fall into supermultiplets of some spacetime
supersymmetry. To see whether these theories are supersymmetric in {\it
curved} space-time the target space supercharges have to be constructed
explicitly by a curved space-time modification of the analysis of
\ref\banks{T. Banks, L.J. Dixon, D. Friedan and E. Martinec, Nucl. Phys. B299
(1988) 613. }.

\subsec{Classification of 4D Heterotic Models}

We can now ask, what other heterotic models can be constructed with the
non-compact group method? Among the cosets listed in \list , the only ones
that lead to models in four curved spacetime dimensions ($D=4$) always include
$SO(d-1,2)/SO(d-1,1)$ for $d\le 4$. The remaining cosets always give models
in more than four dimensions. In this lecture we will concentrate on
four dimensions and therefore use only $SO(d-1,2)$ for $d=2,3,4$. For $D=d=4$
there are no other bosonic coordinates.  When $d\le 3$, then $D-d=4-d$
additional bosonic coordinates are supplied by taking direct products with
other groups (including space-like $U(1)$ or $\IR$ factors) and then gauging
an appropriate subgroup. Furthermore, we include in our list the possibility
of a time-like bosonic coordinate and denote it by a factor of $T$ instead of
$\IR$. All possibilities
 are listed in Table-1 in the column labelled ``right movers".

%table-1

$$\vbox   {    \tabskip=0pt \offinterlineskip
\halign to 370pt  { \vrule# & \strut # &
 #\hfil &\vrule# \ \ &  #\hfil &\vrule# \ \  &  #\hfil &\vrule#
\tabskip =0pt
\cr \noalign {\hrule}
&& \# && left movers with N=1 SUSY && right movers &
\cr\noalign{\hrule}
\noalign {\smallskip}
&&$ 1 $&&${SO(3,2)_{-k}\times SO(3,1)_{1}/ SO(3,1)_{-k+1}} $&
        &$  {SO(3,2)_{-k}/ SO(3,1)_{-k} }  $&
\cr\noalign{\hrule}
&&$ 2 $&&$ {SL(2,\IR)_{-k_1}\times SL(2,\IR)_{-k_2}\times SO(3,1)_1\over
        SL(2,\IR)_{-k_1-k_2+2}}\times \IR $&
        &$  {SL(2,\IR)_{-k_1}\times SL(2,\IR)_{-k_2}\over SL(2,\IR)_{-k_1-k_2}}
 \times \IR   $&
\cr\noalign{\hrule}
&&$ 3 $&&$ \bl (SO(2,2)_{-k}\times SO(3,1)_1/SO(2,1)_{-k+2}\br)\times \IR $&
       &$ \bl (SO(2,2)_{-k}/SO(2,1)_{-k}\br)\times \IR $&
\cr\noalign{\hrule}
&&$ 4 $&&$ SL(2,\IR)_{-k}\times SO(3,1)_1\times \IR $&
        &$ SL(2,\IR)_{-k}\times \IR $&
\cr\noalign{\hrule}
&&$ 5 $&&$ {SL(2,\IR)_{-k_1}\times SL(2,\IR)_{-k_2}\times SO(3,1)_1 \over
          \IR^2 } $&
        &$ {SL(2,\IR)_{-k_1}\times SL(2,\IR)_{-k_2} / \IR^2 }   $&
\cr\noalign{\hrule}
&&$ 6 $&&$ {SL(2,\IR)_{-k_1}\times SU(2)_{k_2}\times SO(3,1)_1 / \IR^2} $&
        &$  {SL(2,\IR)_{-k_1}\times SU(2)_{k_2}/ \IR^2}  $&
\cr\noalign{\hrule}
&&$ 7 $&&$ {(SL(2,\IR)_{-k}\times \IR^2\times SO(3,1)_1)/ \IR }  $&
        &$ {(SL(2,\IR)_{-k}\times \IR^2 )/ \IR }  $&
\cr\noalign{\hrule}
&&$ 8 $&&$ \IR^3\times \IR_Q\times SO(3,1)_1  $&&$ \IR^3\times \IR_Q   $&
\cr\noalign{\hrule}
&& \multispan5 Table-1. Current algebraic description of left movers and
right movers.\hfill &
\cr\noalign{\hrule}
}}$$
For brevity we used $\IR$ where we could have used either $\IR$ or $U(1)$.
Case 3 is obtained from case 2 in the limit $k_1=k_2=k$, while case 4 is the
$k_1=k,\ k_2=\infty$ limit of either case 2 or 5. In case 8, the notation
$\IR_Q$ is used to denote a free boson with background charge $Q$ that
contributes to the central charge $c_Q=1+12Q^2$ just like a Liouville field.
The factors of $\IR$ in cases 2,3 could also be allowed to have a background
charge, but we will assume that it is zero in order to keep the discussion as
simple as possible.

The cases 5,6,7 which contain an $\IR$ factor in the
denominator may further be generalized by multiplying both numerator and
denominator by a factor $\IR^n$. What this implies is that there are many
possible ways of gauging the $\IR$ factors by taking linear
combinations. This may lead to models that have different spacetime dynamics,
however since the central charges remain unchanged, this generalization does
not alter the gauge symmetry results given in Table-2.

The heterotic string will have a supersymmetric left-moving sector and a
non-supersymmetric right-moving sector. The cosets above describe the four
dimensional space-time part of the right-moving sector.  This contributes
$c_R(4D)$ toward the Virasoro central charge. After we analyse the central
charge of the supersymmetric left movers and fix it to be $c_L=15$ in only four
dimensions, we will see that $c_R(4D)$ will be fixed to some value less
than $26$. Therefore, for the mathematical consistency of the theory, we must
require that the right moving sector contains an additional ``internal" part
which makes up for the difference, i.e. $c_R(int)+c_R(4D)=26$. One of the
aims is to compute $c_R(int)$ in each model and then find gauge
symmetry groups that precisely give this value. This procedure will allow us
to discover the gauge symmetries that are possible in these curved spacetime
string models.

To construct a heterotic string we introduce four left moving coset fermions
$\psi^\mu$ that are classified under $H$ as $G/H$ and form a $N=1$
supermultiplet together with the four bosons. The action
that pocesses the superconformal symmetry has the form of $S_2$ in \action\ as
given in \IBSFhet .
The left moving fermions $\psi^\mu$ are coupled to the gauge bosons in $H$. In
the Hamiltonian language, the left moving stress tensor is expressed in the
form of current algebra cosets \KS\IBhe\
as listed in Table-1, where $SO(3,1)_1$ represents the fermions.

This algebraic formulation allows an easy computation of the Virasoro
central charges for left movers $c_L$ as well as the right movers $c_R(4D)$.
For a consistent theory we must set $c_L=15$. This condition puts restrictions
on the various central extensions $k$ and/or background charge $Q$, as
listed in Table-2 (assuming $Q=0$ for cases 2,3,4). After inserting these in
$c_R(4D)$ we find the deficit from the critical value of 26, i.e.
$c_R(int)=26-c_R(4D)$. As seen in the table, the resulting values for
$c_R(int)$ fall within a narrow range. For case 2 or 3 it is possible to
change the central charge within the range $11{1\over 2}<c_R(int) <13$ by
varying  $k_1+k_2$. For the remaining cases it is not possible to
change $c_R(int)$ by using the remaining freedom with the $k's$.

 %%table-2

$$\vbox   {    \tabskip=0pt \offinterlineskip
\halign to 361pt  { \vrule# & \strut # &
 #\hfil &\vrule# \ \ &  #\hfil &\vrule# \ \  &  #\hfil &\vrule#
 \  & #\hfil &\vrule#
\tabskip =0pt
\cr \noalign {\hrule}
&& \# && conditions for $c_L=15 $ && $c_R(int)$ && gauge group, right movers &
\cr\noalign{\hrule}
\noalign {\smallskip}
&&$ 1 $&&$ k=5 $&&$ 11 $&&$ (E_7)_1\times SU(5)_1 $&
\cr\noalign{\hrule}
&&$ 2 $&&$ k_1-2={k_2-2\over 2}(-1+\sqrt {3k_2\over 3k_2-8})   $&
        &$ 13-\delta $&&$ \delta={12\over (k_1+k_2-4)(k_1+k_2-2)}  $&
\cr\noalign{\hrule}
&&$ 3 $&&$ k=3   $&&$ 11{1\over 2} $&
        &$ (E_7)_1\times SU(3)_1\times SU(2)_2\times U(1)_1  $&
\cr\noalign{\hrule}
&&$ 4 $&&$ k=8/3   $&&$ 13  $&&$ (E_8)_1\times SO(10)_1  $&
\cr\noalign{\hrule}
&&$ 5 $&&$ k_1={8k_2-20\over 3k_2-8},\ \ k_1,k_2>{8\over 3}  $&
        &$ 13 $&&$ (E_8)_1\times SO(10)_1 $&
\cr\noalign{\hrule}
&&$ 6 $&&$k_1={8k_2+20\over 3k_2+8}, \ k_2=1,2,3,\cdots $&
        &$ 13  $&&$ (E_8)_1\times SO(10)_1 $&
\cr\noalign{\hrule}
&&$ 7 $&&$ k={8/3} $&&$ 13  $&&$ (E_8)_1\times SO(10)_1  $&
\cr\noalign{\hrule}
&&$ 8 $&&$ Q_0^2={3\over 4} $&&$ 13  $&&$ (E_8)_1\times SO(10)_1  $&
\cr\noalign{\hrule}
&& \multispan7 Table-2. Conditions for $c_L=15$ and examples of symmetries
that give $c_R=26$. &
\cr\noalign{\hrule}
}}$$

The value of $c_R(int)=13$ that occurs for most of the cases is the same as
the deficit for the popular heterotic string models that have four flat
dimensions plus compactified dimensions described by a $c=9$, $N=2$
superconformal theory (i.e. $4+9+13=26$). Hence, for these cases, the
appearance of $(E_8)_1\times SO(10)_1$ as the gauge group has precisely the
same explanation as the usual approach. For the remaining cases we give an
example of a gauge symmetry that will make up the deficit, as listed in Table-
2. Other gauge groups are clearly possible just on the basis of $c_R(int)$;
for example for case 1 see footnote 4.

The gauge symmetry is associated with a conformal theory of right movers. This
additional part of the action may be constructed (as $S_3$ in \action ) from
right moving free fermions with appropriate boundary conditions, or by using
other devices that are quite familiar. We can think of this part as another
current algebra asssociated with the gauge group, and with the central
extensions that are given in Table-2. This final step completes the action for
the model. Further discussion of the model is required to determine the
symmetries consistent with modular invariance. At this stage it is encouraging
to note that {\it the desirable low energy symmetries, including $SU(3)\times
SU(2)\times U(1)$, are contained in these curved space string models that have
only four dimensions}.

\newsec{Geometry of the Manifold}

A gauged WZW model can be rewritten in the form of a non-linear sigma model by
choosing a unitary gauge that eliminates some of the degrees of freedom from
the group element, and then integrating out the non-propagating gauge fields
\BN\WIT . The remaining degrees of freedom are identified with the string
coordinates $X^\mu(\tau,\sigma)$.
The resulting action exhibits a gravitational metric $G_{\mu\nu}(X)$ and an
antisymmetric tensor $B_{\mu\nu}(X)$ at the classical level. At the one loop
level there is also a dilaton $\Phi(X)$. These fields govern the spacetime
geometry of the manifold on which the string propagates. Conformal invariance
at one loop level demands that they satisfy coupled Einstein's equations.
Thanks to the exact conformal properties of the model these equations are
automatically satisfied. Therefore, any of our non-compact gauged WZW models
can be viewed as generating automatically a solution of these rather
unyielding equations. One only needs to do some straightforward algebra to
extract the explicit forms of $G_{\mu\nu}, B_{\mu\nu}, \Phi$.

This algebra can be carried out by starting from the Lagrangian, such as in
\action , and has been done for all the models in four dimensions listed in
Table-1. The first case was $SL(2,\IR)/\IR$ which was interpreted by Witten
\WIT\ as the geometry of a 2D black hole. The higher dimensional cases yield
more intricate but singular geometries \IBSFthree\others\IBSFhet\GIN\ .
Although the Lagrangian method is straightforward, it has a number of
drawbacks. First, it yields the geometry only in a patch that is closely
connected to a particular choice of a unitary gauge. The ramaining patches of
the global geometry can be recovered only in other unitary gauges and may have
no resemblance to the analytic form of the metric, dilaton, etc. in another
unitary gauge. To overcome this problem we have introduced global coordinates
\IBSFglobal\ on the complete geometry. The global coordinates are gauge
invariant. The second problem with the Lagrangian method is that it yields the
semi-classical geometry up to one loop in an expansion in powers of $1/k$.
However, since the gauged WZW model is conformally exact one would rather
obtain the conformally exact geometry by using alternative methods. It turns
out that the Hamiltonian method that utilizes the GKO construction solves both
of these problems simultaneously and yields an exact metric and dilaton to all
orders in $1/k$ \IBSFexa . Therefore in this lecture we concentrate on the
Hamiltonian approach.

With the Hamiltonian approach one can compute the
gravitational metric and dilaton backgrounds to all orders in the quantum
theory (all orders in the central extension $k$) at the ``classical" level
(i.e. no string loops). We have managed to obtain these quantities for
bosonic, type-II supersymmetric, and heterotic string theories in $d\le 4$.
It turns out that the geometry of the heterotic and type-II superstrings are
obtained by deforming the geometry of the purely bosonic string by definite
shifts in the exact $k$-dependence. Therefore, it is sufficient to first
concentrate on the purely bosonic string.
The following relations have been proven for $G/H=SO(d-1,2)/SO(d-1,1)$ which
is relevant to string theory \IBSFexa : (i) For type-II superstrings the
conformally {\it exact} metric and dilaton are identical to those of the
non-supersymmetric {\it semi-classical} bosonic model except for an overall
renormalization of the metric obtained by $k\to k-g$. (ii) The exact
expressions for the heterotic superstring are derived from their exact bosonic
string counterparts by shifting the central extension $k\to 2k-h$ (but an
overall factor  $(k-g)$ remains unshifted). (iii) The combination
$e^\Phi\sqrt{-G}$ is independent of $k$ and therefore can be computed in
lowest order perturbation theory. Cases 2,5,6 in Table-1 are a
bit more complicated because of the two central extensions, but the results
that relate the bosonic string to superstrings are analogous. Case 6 is
explicitly discussed in \IBSFslsu , and the others are just analytic
continuations of this one.

The main idea is the following. For the bosonic string the conformally exact
Hamitonian is the sum of left and right Virasoro generators $L^L_0+L_0^R$.
They may be written purely in terms of Casimir operators of $G$ and $H$ when
acting on a state $T(X)$ at the tachyon level. The exact dependence on the
central extension $k$ is included in this form by using the GKO formalism in
terms of currents. For example for the left-movers
 \foot{ Here $J^L_G$ and $J^L_H$ are {\it antihermitian} group and subgroup
generators obeying the appropriate Lie algebras, and $g$, $h$ are the Coxeter
numbers for the group and the subgroup. For the cases of interest in this
paper $g=d-1$, $h=d-2$ for $d\ge 3$, and $g=2,\ h=0$ for $d=2$. }

\eqn\lzer{\eqalign{&L^L_0 T=\bl({\D^L_G\over k-g}-{\D^L_H\over k-h}\br)T\cr
&\D^L_G \equiv Tr(J^L_G)^2, \qq \D^L_H \equiv Tr(J^L_H)^2\ ,\cr} }
The exact quantum eigenstate $T(X)=<X|T>$ can be analyzed in $X$-space. Then
the Casimir operators become Laplacians constructed as differential operators
in group parameter space ($dim G$). Consider a state $T(X)$ which is a singlet
under the gauge group $H$ (acting simultaneously on left and right movers)

\eqn\cond{(J^L_H +J^R_H)\ T=0\ .}

\no
Because of the $dimH$ conditions $T(X)$ can depend only on
$d=dim(G/H)$ parameters, $X^a$ (string coordinates), which are
$H$-invariants constructed from group parameters (see below). The fact that
there are exactly $dim(G/H)$ such independent invariants is not immediately
obvious but it should become apparent to the reader by considering a few
specific examples. As discussed in \IBSFglobal\ these are in fact the
coordinates that globally describe the sigma model geometry. Consequently,
using the chain rule, we reduce the derivatives in \lzer\ to only derivatives
with respect to the $d$ string coordinates $X^a$. In this way we can write the
conformally exact Hamiltonian $L_0^L+L_0^R$ as a Laplacian differential
operator in the global curved space-time manifold involving only the string
coordinates $X^a$. By comparing to the expected general form

\eqn\laplacian{ (L_0^L+L_0^R)T={-1\over e^\Phi\sqrt{-
G}}\partial_a(e^\Phi\sqrt{-G}G^{ab}\partial_bT)}
for the singlet $T$, we read off the exact global metric and dilaton.

We have applied this program to all the models in Table-1 and obtained the
exact geometry to all orders in $1/k$. The large $k$ limit of our results
agree with the semi-classical computations of the Lagrangian method.
In the special case of two dimensions we also agree with another previous
derivation of the exact metric and dilaton for the $SL(2,\IR)/\IR$ bosonic
string
 \ref\DVV{R. Dijgraaf, E. Verlinde, H. Verlinde, Nucl. Phys. {\bf B371} (1992)
269.}.
 We summarize here the global and conformally exact results for the metric and
dilaton in the case of $SO(d-1,2)_{-k}/SO(d-1,1)_{-k}$ for d=2,3,4 \IBSFexa .
Due to the more complex expressions we refer the reader to the original
literature for the remaining cases
 \ref\SF{K. Sfetsos, ``Conformally Exact Results for for $SL(2,\IR)\times
SO(1,1)^{d-2}/SO(1,1)$ Coset Models", USC-92/HEP-S1 (hep-th/9206048), to
appear in Nucl. Phys. B }
\IBSFslsu . The group element $g$ for
$SO(d-1,2)/SO(d-1,1)$ can be parametrized as a $(d+1)\times (d+1)$ matrix in
the form

\eqn\group { g=\left ( \matrix {1 & 0 \cr 0 & ({1+a\over 1-a})_{\m}{}^{\n}\cr }
 \right  )  \ \left (\matrix {b  & (b+1) x^\nu \cr
-(b+1) x_\mu  & (\eta_\mu^{\ \nu} -(b+1) x_\mu x^\nu) \cr } \right ),}

\no
where $b={1-x^2\over {1+x^2}}$. The $d$ parameters $x_\mu$ and $d(d-1)/2$
parameters $a_{\mu\nu}$ transform as vector and antisymmetric tensor
respectively under the Lorentz subgroup $H=SO(d-1,1)$ which acts on both sides
of the matrix as $g\rightarrow hgh^{-1}$. By considering the infinitesimal
left transformations $\d_L g=\e_L g$ we can read off the
generators that form an $SO(d-1,2)$ algebra for left transformations.

\eqn\leftgen{\eqalign{&J^L_{\m\n}={1\over 2}(1+a)_{\m\a}(1+a)_{\n\b}
{\partial\over \partial a_{\a\b}}\cr
 &J^L_{\m}=-{1\over 2}(1+x^2)\bl({1+a\over 1-a}\br)_{\m}{}^{\n}
{\partial \over
\partial x^{\n}} +{1\over 2}(1+a)_{\m\a}(1+a)_{\b\g}x^{\g}
{\partial\over \partial a_{\a\b}}\ .}  }

\no
If we consider instead, the infinitesimal right transformations
$\d_R g=g \e_R$ we find the following expressions for the generators of right
transformations

\eqn\rightgen{\eqalign{&J^R_{\m\n}=
-{1\over 2}(1-a)_{\m\a}(1-a)_{\n\b}
{\partial\over\partial a_{\a\b}}-x_{[\m}{\partial \over \partial x^{\n]}}\cr
&J^R_{\m}={1\over 2}(x^2-1){\partial\over \partial x^{\m}}-x_{\m}x^{\n}
{\partial \over \partial x^{\n}}
-{1\over 2}(1-a)_{\m\a}(1-a)_{\g\b}x^{\g}
{\partial\over \partial a_{\a\b}}\ .}}

\no
The $J^R$ currents obey the same commutation rules as $J^L$ and moreover
commute with each other $[J^L,J^R]=0$.  The quadratic Casimirs for the
group and subgroup on either the left or the right are obtained by squaring
these currents. For the explicit expressions see \IBSFexa .

As argued above the global parametrization of the manifold is given in terms
of H-invariants, i.e. Lorentz invariants in the present case. In order to
obtain a diagonal metric on the manifold one must find $d$ convenient
combinations of these Lorentz invariants in $d$ dimensions. We give here the
basis that diagonalizes the semi-classical metric at large $k$. One of the
natural invariants already occurs in the construction of the group element for
every $d$, namely $b={1-x^2\over 1+x^2}$.

\subsec{Two dimensions}

For $d=2$ the antisymmetric tensor is Lorentz invariant
$a_{\mu\nu}=a\epsilon_{\mu\nu}$, and it is convenient to parametrize
$a=tanh(t)\ or\ coth(t)$. Then the global string coordinates can be taken as
$X^a=(t,b)$. Given all possible values for $(a,x^\mu)$ the ranges of the two
invariants cover the entire plane $-\infty<t,b<+\infty$.
The metric is given by the line element

\eqn\twometric{ds^2=2(k-2)\bl({db^2\over 4(b^2-1)}
-\b(b) {b-1\over b+1} dt^2\br),
\qq \b^{-1}(b)=1-{2\over k} {b-1\over b+1}\ . }

\no
For the dilaton the corresponding expression is

\eqn\twodilaton{\Phi=\ln\bl({b+1\over \sqrt{\b(b)}}\br)+const\ . }

\no
The scalar curvature for this metric is

\eqn\curv{R={2k\over k-2} {(k-2)b+k-4\over \bl((k-2)b+k+2\br)^2}\ .}

\no
The curvature is singular at $b=-(k+2)/(k-2)$, which is also where
$\beta(b)=\infty$. These are the properties of the exact 2d metric. The semi-
classical metric is obtained by taking the large $k$ limit, for which $\b=1$.
Then the singularity is at $b=-1$. Following Witten this singularity is
interpreted as a black hole while the horizon is at $b=1$. The signature of
the space is $(+-)$ or $(-+)$ depending on the region in the $(t,b)$ plane as
indicated in Fig-2 of \IBSFglobal . The signature is understood by examining
the semi-classical metric. To see the connection to the Kruskal coordinates
used by Witten let $b=1-2uv$ and $u^2=e^{2t}|b-1|/2$, $v^2=e^{-2t}|b-1|/2$.

There are asymptotically flat  regions which are displayed by the change of
coordinates $b=\pm\cosh {2z_1\over\sqrt{2(k-2)}},\ t={z_0\over \sqrt{2k} }$.
For large $z_1\to\pm\infty$ and any $z_0$ the exact metric and dilaton have
the asymptotic forms

\eqn\asymtwo{ ds^2=dz_1^2-dz_0^2,\qquad \Phi=\sqrt{2\over k-2}|z_1| ,}
displaying a dilaton which is asymptotically linear in the space direction,
just like a Liouville field in 2d quantum gravity with a background charge.
Despite the flat metric there is no Poincar\'e invariance due to the linear
dilaton. Note that both the region outside the horizon ($b\to +\infty$) and
the naked singularity region ($b\to -\infty$) are asymptoticaly flat.

\subsec {Three dimensions}

For $d=3$ the antisymmetric tensor is equivalent to a
pseudo-vector $a_{\mu\nu}=\epsilon_{\mu\nu\lambda}y^\lambda$, from which we
construct two convenient invariants $v=2/(1+y^2)$ and $u=-v(x\cdot y)^2/x^2$,
which together with $b$ provide a basis for the string coordinates
$X^a=(v,u,b)$. Given all possible values taken by $(x^\mu,y^\mu)$ the allowed
ranges for the invariants are

\eqn\rangethree{ \eqalign {(+-+)\ or\  (-++)\quad & \{b^2>1 \ \& \ uv>0\}, \cr
(++-)\quad &\{b^2<1\ \& \ uv<0\}, \quad except \quad 0<v<u+2<2 . } }
The 3d conformally exact metric is given by the line element \IBSFexa

\eqn\trimetric{ds^2=2(k-2)\bl(G_{bb} db^2 +G_{vv} dv^2 +G_{uu} du^2
+2G_{vu} dvdu\br)\ . }
where
\eqn\tridef{\eqalign{&G_{bb}={1\over 4(b^2-1)}\cr
&G_{vv}=-{\b(v,u,b)\over 4v(v-u-2)}
\bl({b+1\over b-1} + {1\over k-1} {u+2\over v-u-2}\br)\cr
&G_{uu}={\b(v,u,b)\over 4u(v-u-2)}
\bl({b-1\over b+1} - {1\over k-1} {v-2\over v-u-2}\br)\cr
&G_{vu}={1\over 4(k-1)} {\b(v,u,b)\over (v-u-2)^2}\ ,\cr}  }
and

\eqn\defb{\b^{-1}(v,u,b)=1+{1\over k-1} {1\over v-u-2}
\bl({b-1\over b+1} (u+2) -{b+1 \over b-1} (v-2) -{2\over k-1}\br)\ . }
The exact dilaton is

\eqn\tridilaton{{\Phi}=\ln \bl({ (b^2-1) (v-u-2)\over \sqrt{\b(v,u,b)}}\br)
+ \Phi_0\ , }

\no
In the large $k$ limit one obtains the global version of a semi-classical
metric derived in \IBSFglobal\ with Lagrangian methods

\eqn\semicl{ {ds^2\over 2(k-2)}{\big\vert}_{k\to \infty}={db^2\over 4(b^2-1)}-
{1\over v-u-2}\bl({b+1\over b-1} {dv^2\over 4v}-{b-1\over b+1}{du^2\over 4u}\br
) }
The signature $(+-+),\ or \ (-++),\ or\ (++-)$ depends on the region and is
indicated in Fig-1 of \IBSFglobal . A three dimensional view of this metric is
given in Figs-4 of \IBSFglobal . The surface is where the scalar curvature
blows up. This coincides with the location where the dilaton blows up in the
large $k$ limit as seen from the above expression. The space has two
topological sectors denoted by the sign of a conserved ``charge"
$\pm=sign(v(b+1))=sign(u(b-1))$. The sign never changes along geodesics. A
more intuitive view of the space is obtained in another set of coordinates for
the plus sector $(b,\lambda_+,\sigma_+)$ and the minus sector $(b,\lambda_-
,\sigma_-)$, which are given by $\lambda^2_\pm=\pm v(b+1)$ and
$\sigma^2_\pm=\pm u(b-1)$. Then the singularity surface is shown in Figs-3 of
\IBSFglobal .
In the plus region the singularity surface has the topology of the double
trousers with pinches in the legs. In the minus region we have the topology of
two sheets that divide the space into three regions.

There are asymptotically flat regions that may be displayed by a change of
variables to $b=\pm \cosh {1\over \sqrt{3(k-2)}}(2z_1-z_0)$,
$u=(\pm)\prime\cosh
 {1\over \sqrt{3(k-2)}}(-z_1+2z_0) \cosh^2z_2$, $v=(\pm)\prime \cosh {1\over
\sqrt{3(k-2)}}(-z_1+2z_0) \sinh^2z_2$. For large values of $z_1\to\pm\infty$,
and finite values of $(z_0,z_2)$, the semiclassical metric and dilaton take
the form

\eqn\asymthree{ ds^2=-dz_0^2+dz_1^2+dz_2^2, \qquad
\Phi=\sqrt{6\over k-2}|z_1'|,}
showing that the dilaton is linear in a space-like direction
$z_1'={5\over 3}z_1-{4\over 3}z_0$ in the
asymptotically flat region. Then $z_1'$ behaves just like a Liouville field.
The asymptotic metric may be written as $ds^2=-(dz_0')^2+(dz_1')^2+dz_2^2$
in terms of the Lorentz transformed $z_1'$ and $z_0'={5\over 3}z_0-{4\over 3}
z_1$. The exact metric is not flat when only $|z_1|$ is large. To display its
asymptotically flat region one requires somewhat different coordinates.

\subsec{Four dimensions}

For $d=4$ one can construct the Lorentz invariants

\eqn\testates{x^2\ ,\qq z_1={1\over 4}Tr(a^2)\ ,\qq
z_2={1\over 4}Tr(a^* a)\ , \qq z_3=xa^2x/x^2 \ ,}

\no
where $a^*_{\m\n}={1\over 2}\e_{\m\n\a\b}a^{\a\b}$ is the dual of $a_{\m\n}$.
However, the semi-classical metric  is diagonal for a different set of four
invariants $X^a=(v,u,w,b)$ given by

\eqn\relat{\eqalign{&b={1-x^2 \over 1+x^2}\ , \qq
u={1+z_2^2+2(z_1 -z_3) \over 1-2 z_1-z_2^2}\cr
&v={1+z_1 +\sqrt{z_1^2 +z_2^2} \over 1-z_1 -\sqrt{z_1^2 +z_2^2}}\ ,\qq
w={1+z_1 -\sqrt{z_1^2 +z_2^2} \over 1-z_1 +\sqrt{z_1^2 +z_2^2}}\ .} }

\no
To find the ranges in which the above global coordinates take their values we
consider a Lorentz frame that can cover all possibilities without loss of
generality. First we notice that by Lorentz transformations the antisymmetric
matrix $a_{\m\n}$ can always be transformed to a block diagonal
matrix with the non-zero elements

\eqn\blo{a_{01}=\tanh t\ {\rm or}\ \coth t\ ,\qq a_{23}=\tan \phi\ .}

\no
Then using \relat\ one can deduce the form of the global variables:
$v=\pm \cosh 2t$, $w=\cos 2\phi$, and $u={1\over x^2}
\bl(w(x_0^2-x_1^2)-v(x_2^2 +x_3^2)\br)$. Therefore the string variables
can take values in the following regions with the signature in the
$(v,u,w,b)$ basis

\eqn\range{\eqalign{
& (-+++):\quad b^2>1,\quad \{-1<w<u<1<v\ \ or \ \ v<-1<u<w<1 \cr
& \hskip 5cm or\ -1<w<1<u<v \},\cr
& (+-++):\quad b^2>1, \quad \{-1<w<1<v<u\ \ or \ \ u<v<-1<w<1 \}\cr
& (+++-):\quad b^2<1, \quad \{u<w<11<v \ or \ v<-1<w<u \ or \ v<u<-1<w<1\}. }}

With this set of coordinates we compute the conformally exact dilaton and
metric as before. The dilaton field is

\eqn\fourdilaton{\Phi=\ln \bl({(b^2-1)(b-1)(v-u)(w-u)\over
\sqrt{\b(b,u,v,w)}}\br) +\Phi_0\ . }
and the metric is given by

\eqn\fourmetric{\eqalign{ds^2=2(k-3)
&\bl(G_{bb} db^2 +G_{uu} du^2 +G_{vv} dv^2 +G_{ww} dw^2\cr
&+2 G_{uv} du dv +2 G_{uw} du dw +2 G_{vw} dv dw\bl)\ ,\cr} }
where

\eqn\fourmet{\eqalign{&G_{bb}={1\over 4(b^2-1)}\cr
&G_{uu}={\b(b,u,v,w)\over 4(u-w)(v-u)}\biggl({b-1\over b+1}-{1\over k-2}
{(v-w)^2\over {(v-u)(u-w)}} (1-{1\over k-2} {b+1\over b-1})\biggr)\cr
&G_{vv}=-{(v-w)\b(b,u,v,w)\over 4(v^2-1)(v-u)}\biggl({b+1\over b-1}-{1\over k-
2} {1\over (v-u)(u-w)}\bl[1-u^2+\cr
 &\quad +({b+1\over b-1})^2 (v-u)(v-w) +{1\over k-2}
{b+1\over b-1} {(1+v^2)(u+w)-2v(1+uw)\over v-w}\br]\biggr)\cr
&G_{ww}={(v-w)\b(b,u,v,w)\over 4(1-w^2)(u-w)}\biggl({b+1\over b-1}-{1\over k-2}
{1\over (v-u)(u-w)}\bl[1-u^2+\cr
 &\quad +({b+1\over b-1})^2 (u-w)(v-w)-{1\over k-2}
{b+1\over b-1} {(1+w^2)(u+v)-2w(1+uv)\over v-w}\br]\biggr)\cr
&G_{uv}={\b(b,u,v,w)\over 4(k-2)(v-u)^2}\bl(1-{1\over k-2} {b+1\over b-1}
{v-w\over u-w}\br)\cr
&G_{uw}={\b(b,u,v,w)\over 4(k-2)(u-w)^2}\bl(1-{1\over k-2} {b+1\over b-1}
{v-w\over v-u}\br)\cr
&G_{vw}={1\over (k-2)^2} {b+1\over b-1} {\b(b,u,v,w)\over
4(v-u)(u-w)}\ ,\cr} }

\no
and the function $\b(b,u,v,w)$ is defined by

\eqn\defbb{\eqalign{ & \b^{-1}(b,u,v,w)=1+{1\over k-2} {(v-w)^2
\over (v-u)(w-u)} \biggl({b+1\over b-1}+{b-1\over b+1} {1-u^2\over (v-w)^2}\cr
&\quad +{1\over k-2}\bl({vw+u(v+w)-3 \over (v-w)^2} -({b+1\over b-
1})^2\br)\biggr) +{2\over (k-2)^3} {b+1\over b-1} {vw-1 \over (v-u)(u-w)}\
.\cr} }

The large $k$ limit of these expressions reduce to the semiclassical dilaton
and metric that follow from the Lagrangian approach

\eqn\semetr{\eqalign{{ds^2\over 2(k-2)} {\big \vert}_{k\to \infty}=
&{db^2\over 4(b^2-1)}+{b-1\over b+1} {du^2\over 4(v-u)(u-w)}\cr
&+{b+1\over b-1}(v-w)\bl({dw^2\over 4(1-w^2)(u-w)}
-{dv^2\over 4(v^2-1)(v-u)}\br)\ .\cr} }

We can see that the signature of the semiclassical metric for different ranges
of the parameters \range\ is precisely as required by the group parameter
space which led to \range . However, for the exact metric $\beta(u,v,w,b)$ must
remain positive to keep $-det(G)$ positive. This implies that part of the
regions in \range\ are screened out by quantum effects for the exact geometry.
This screening phenomenon is true for every dimension $d=2,3,4$ and the
screened regions must be interpreted in the quantum theory as tunneling or
decay regions for probability amplitudes (such as the tachyon wavefunction).
Under any circumstances the manifold cannot go outside of the range \range\
dictated by the group theory.

As in the previous $d=2,3$ cases, we can check that our explicit expressions
for the dilaton and metric give the $k$-independent combination $\sqrt{-
G}e^{\Phi}$. Therefore this quantity takes the same value for either the
exact metric and dilaton or the semiclassical metric and dilaton. Since it
is unrenormalized by quantum effects (other than one loop), it may be
computed in lowest order perturbation theory. This combination appears in the
Dalambertian and is also closely related to the integration measure in the
path integral. Through group theoretical arguments given in
\IBSFthree\IBSFhet\ it was possible to guess that this combination should
remain unrenormalized by quantum effects
 \foot{In previous papers \IBSFthree\IBSFhet\ we erroneously stated that the
path integral for the gauged WZW model needed an additional factor
$F=e^{\Phi}/\sqrt{-G}$. This point was a mistake because our argument did
not take an additional anomaly factor into account. I thank A. Tsyetlin for
discussions on this issue. The correct measure for the gauged WZW model is
just the Haar measure for the group element times the naive measure for the
gauge fields. In a unitary gauge that reduces the action to a sigma model,
the integration over the gauge fields produces a determinant and an anomaly
factor such that, when combined with the Haar measure and a Faddeev-Popov
determinant, the effective measure takes the form of the volume element in
curved space, i.e. $D^dX\sqrt{-G}$, which is the expected result
 \ref\volume{R. Jackiw and S. Weinberg, Phys. Rev. {\bf D3} (1971) 2486. }. }.
 So far there has not been a more satisfactory explanation of the
non-renormalization of this quantity.

Similar to the $d=2,3$ cases the 4d manifold has an asymptotically flat
region, but it will not be discussed here.

\subsec{Particle and String Geodesics}

Having global coordinates and a global geometry is not sufficient to get a
feeling of the geometry, one also needs to know the behavior of the
geodesics. However, for the complicated metrics that are displayed above the
geodesic equation seems to be completely unmanageable. Fortunately, we have
developed a procedure that relies on group theory and managed to solve for all
particle geodesics. The trick is to take advantage of the fact that the global
coordinates are gauge invariant under $H$-transformations. Then we may solve
the equations of motion for the group element $g$ in any gauge, and use
{\it the solution} to construct the $H$-invariant combinations that form the
global coordinates of the geometry. In fact, there is an axial gauge in which
$g$ is solved easily \IBSFglobal . For a point particle (string shrunk to a
point) it is given as a function of proper time

\eqn\groupel{ g(\tau)=e^{\alpha\tau} g_0 e^{(p-\alpha)\tau}, }
where $g_0$ is a constant group element at initial proper time $\tau$, and
$\alpha , p$ are constant matrices in the Lie algebras of $H$ and $G/H$
respectively. The equations of motion require that these constants satisfy a
constraint

\eqn\constraint{ (g_0(p-\alpha)g_0^{-1})_H+\alpha=0\ ,}
where the subscript $H$ implies a projection to the Lie algebra of $H$. This
solution applies to any group and subgroup. As shown in \IBSFglobal\ the
standard geodesics equations for the geometries displayed above are
automatically solved when the $H$-invariants are constructed from the solution
\groupel\constraint . In this way all light-like, space-like and time-like
geodesic solutions are obtained.

With the point geodesics at hand we have learned a number of additional
interesting properties about the $d=2,3,4$ manifolds \IBSFglobal\ which
generalize to other non-compact gauged WZW models as well. The most striking
feature is that the manifolds that are pictured in the figures have many
copies and the complete manifold must include all the copies. The gauge
invariant coordinates (e.g. $(b,t)$ for $d=2$) are not sufficient to fully
describe the structure. There are additional {\it discrete} gauge invariants
constructed from the group element $g$ that label the copies of the manifold.
This can be seen easily in our examples since the gauge subgroup is just the
Lorentz group and its properties are well known. In this case the invariants
are Lorentz dot products constructed from a vector $x^\mu$ and a tensor
$a^{\mu\nu}$. Let us consider the invariant $b=(1-x^2)/(1+x^2)$, say in the
region $x^2>0$. It is known that the time component $x^0$ could be either
positive or negative and that a Lorentz transformation cannot change this
sign. Therefore, the sign of $x^0$ is a discrete gauge invariant which does
not show up in the metric or dilaton that characterized the manifolds
discussed above. However, the model as whole knows about this sign through the
group element $g$. Such discrete invariants are present in every
{\it non-compact} gauged WZW model and they label copies of the manifolds
described above. We may then ask whether these copies communicate with each
other? The answer is yes, they do, and this can be seen by following the
behaviour of a particle geodesic. The full information about the particle
geodesic is contained in the solution for $g$ in \groupel\constraint . From
this it can be verified that at the proper time that a particle touches a
curvature singularity the discrete invariant switches sign and then the
particle continues its journey smoothly from one copy of the manifold to the
next. For example, in the 2d black hole case this happens for a time-like
geodesic (i.e. massive particle) in a finite amount of proper time (on the
other hand, a light-like geodesic takes an infinite amount of proper time to
reach the singularity and therefore ends its journey without changing copies
of the manifold). This behavior is present in all non-compact models in this
paper as well as other models (e.g. we have verified it in the $
SL(2,\IR)\times SU(2)/\IR^2 $ model). It is reminiscent of the
Reissner-Nordtsrom black hole in which geodesics move on to other worlds. The
difference is that in our case this happens at the singularity itself. When
quantum corrections are included and the exact metric considered, then the
singularity and the transition to other worlds no longer seem to be at the
same place, at least this is the case for the 2d black hole.
The spectrum of the discrete invariant depends on the group
representation and therefore one expects different numbers of copies in
different quantum states. The number of copies is infinite for quantum states
with non-fractional quantum numbers, which is typical in unitary
non-holomorphic representations of non-compact groups. When the number of
copies is infinite the particle can never come back to the same world, but
for a finite number of copies the particle returns to the original world by
emerging from a white singularity.

So far we have discussed particle geodesics that correspond to a string
collapsed to a single point. We may also investigate string geodesics in the
same manifolds. That is we are also interested in solutions for the strings
moving in curved spacetime, just like one has a complete solution in flat
spacetime in terms of harmonic oscillator normal modes. This problem has been
solved in principle for the non-compact gauged WZW models in \IBSFthree .
There the solution for the group element $g(\tau,\sigma)$ has been obtained
explicitly in terms of normal modes. This is the analog of \groupel\ above.
There remains to construct the appropriate dot products to form the
invariants, which in turn are the solutions to the string geodesics. This last
part has not yet been performed explicitly, but it is only a matter of
straightforward algebra of the kind performed for the particle geodesics in
\IBSFglobal . This procedure gives all the solutions in curved spacetime and
can answer questions of the type ``what happens when a string falls into a
black hole ?"

\subsec{Duality}

Due to the lack of space we have not covered other interesting topics such as
duality properties of these manifolds. It was shown in \IBSFthree\IBSFglobal\
that there is a dynamical duality that generalizes the $R\rightarrow 1/R$
duality properties of conformal field theories based on tori. This is related
to the axial/vector duality that is present in the 2d black hole. It was shown
in \IBSFglobal\ that the duality transformation is equivalent to an inversion
in group parameter space $(x_\mu , a_{\mu\nu})$ given in \group . This
inversion generates discrete leaps for the group parameter that corresponds to
interchanging different regions of the geometrical manifold. For details the
reader is refered to \IBSFglobal . This duality property is closely related to
mirror symmetry of the kind discussed for Calabi-Yau manifolds, as will be
explained elsewhere. The duality symmetry mentioned here is different than the
one discussed in recent months by Verlinde, Giveon, Rocek and others.

\newsec{Some Open problems}

The special property of the models constructed in this lecture is that they can
be further investigated by using current algebra techniques. The simplest
model is case 8, since it is essentially flat, its quantum theory reduces
to the manipulation of harmonic oscillators. For the remaining models the
spectrum of low energy particles is obtained by computing the quadratic
Casimir operators of the non-compact groups that define the model. The
computation of the spectrum will be reported in a future publication. Since
the flavor groups such as $SU(3)\times SU(2)\times U(1)$ or $SU(5),\ SO(10)$,
etc. appear at level 1, it is already evident that the quark and lepton type
of matter will appear in color triplets and singlets and SU(2) doublets and
singlets.

To compute the spectrum one needs to know all the representations of the
non-compact groups that appear in Table-1. An inspection of the table shows
that one needs the following representations

\item {(i)} $SL(2,\IR)$ in the basis in which one of the non-compact
generators is diagonal. This basis is labelled by $|j\mu>$, the range of the
allowed values of $j$ are known in a unitary representation, but the
literature is not too clear on the allowed values of $\mu$.

\item {(ii)} $SL(2,\IR)\times SL(2,\IR)$ in the basis in which the diagonal
subgroup $SL(2,\IR)$ labels the states. This basis has the form
$|j_1,j_2;j,m>$, just like in the problem of addition of angular momentum. The
allowed values of $j_1,j_2$ in unitary representations of $SL(2,\IR)\times
SL(2,\IR)$ are the standard ones. For given $j_1,j_2$ the allowed values of
$j$ are not fully classified.

\item {(iii)} $SO(3,2)$ in the basis in which $SO(3,1)$ is diagonal. This
basis has the form $|a,b;c,k;j,m>$ where $(a,b)$ label the quadratic and
quartic Casimirs of $SO(3,2)$, $(c,k)$ label the two Casimirs of the Lorentz
group $SO(3,1)$ and $(j,m)$ labels the $SO(3)$ rotation subgroup. In unitary
representations the allowed ranges for each one of these pairs are known,
however, for given $(a,b)$ it is not generally known what would be the allowed
values of $(c,k)$.

Evidently there is some mathematics to be developed to
obtain the full spectrum. After determining the required ranges of quantum
numbers, the strategy is to determine those states that satisfy the conformal
conditions for the Virasoro generators in the Neveu-Schwarz and Ramond sectors
of the theory (i.e. $L_0=1$ in the bosonic sector, etc.). These states will
include particles of various spins, chirality and gauge quantum numbers. The
ones that are relevant to low energy physics are those protected from getting
masses by chiral invariance and gauge invariance. In particular the spin one
bosons will fall into the adjoint represention of the gauge groups in Table-2,
while the chiral fermions will be the candidates for quarks and leptons. These
matter multiplets are expected to appear in the fundamental representation of
the gauge group when the Kac-Moody level is exactly one. One of the
intersting questions is the number of families that will emerge in this
computation. I speculate that the number of distinct non-compact group
representations that satisfy the same eigenvalue condition for $L_0$ may be
interpreted as the number of families. This may have some relation to the
geometrical or topological or duality properties of the manifold. This program
for computing the spectrum is underway.

We have only scratched the surface of the subject of non-compact gauged WZW
models. We have shown that this approach is very useful for learning about
strings in curved spacetime that may be relevant for the early part of the
Universe. It is during this era that string theory should be relevant and it is
during this era that the matter we know was formed. Therefore, in trying to
solve the puzzles of the Standard Model with respect to the spectrum of matter
and gauge bosons we may hope that a string theory in curved spacetime may
guide us. For this reason I believe that it is valuable to study in great
detail the models presented in Table-1. These are solvable models that should
direct us toward a realistic unified theory.

%%%%%%%%%%%%%%%%%%%%%%%%%%%%%%%%%%%%%%%%%%%%%%%%%%%%%%%%%%%%%%

\listrefs
\end